# Near-Field Heat Transfer Percolation

# in Nanoparticles based Composite Media

Sebastian Volz[1] and Gilberto Domingues[2]

[1] Laboratoire d'Energétique Moléculaire et Macroscopique, Combustion, UPR CNRS 288

Ecole Centrale Paris, 92295 Châtenay Malabry, France

Email: volz@em2c.ecp.fr

T: 33 1 4113 1049, F: 33 1 4702 8035

[2] Laboratoire Microstructrures et Comportements, CEA/Le Ripault, BP 16 - 37260 Monts, France

**Abstract**

Near-field radiative heat transfer is investigated in composite media including nanoparticles. By modeling pair interactions only, the effective thermal conductivity due to near field radiation is calculated based on a thermal nodes model. We highlight the onset of a Near-Field percolation occurring much earlier than the mechanical percolation at critical volume fraction f=0.033. This mechanism drastically increases the thermal conductivity even at low volume fractions. It also indicates a simple experimental protocol to prove Near-Field contact.

## 1. Introduction

The enhancement of the thermal conductivity of thermal insulators such as cooling liquids, glues, and electrical insulators is a key issue in the fields of heat exchangers and of microelectronics. Embedding nano-objects with a high thermal conductivity is an efficient solution in the case fluids[1] although the physical explanations for the unexpected high thermal conductivity increase are not clear yet. Efficient heat transfer in the nano-object distribution requires the formation of the percolation network. Accordingly, the effective thermal conductivity $\lambda$ should depend on the critical volume fraction $f_c$ as follows:[2] $\lambda=\lambda_0 (f-f_c)^t$ where $\lambda_0$ is the thermal conductivity of the nano-object. Recent works[3] proposed to use nano-cylinders or tubes because the critical volume fraction is reverse proportional to the aspect ratio.[2] The cylinder/matrix thermal resistance however limits the thermal conductivity.[4]

We recently have presented the Molecular Dynamics (MD) calculation of the heat flux between silica nanoparticles due to near-field interaction.[5] When the separation distance d is larger than two nanoparticles (NP) diameters, the dipole-dipole approximation fits the MD results and predict the well-known $d^{-6}$ dependence. When d is lower than two NP diameters, the dipole-dipole model is not relevant anymore and MD reveals a drastic increase of the thermal conductance. A power law of about –20 is obtained in this range. In those conditions, the volume fraction is lower than 0.07 which makes thermal transfer through near-field radiation most probable in nanocomposites with adequate volume fractions.

The percolation threshold is due to a discontinuity in the interaction between elements. For instance, the mechanical contact that occurs very abruptly, generates clear discontinuities in the electrical and thermal conduction properties. Near field radiation is a continuous interaction that should not yield the same behaviour. However, the strong variation of the interaction when d=4a is likely to generate a threshold related to 'near field contact' between particles. This macroscopic effect could provide a clear experimental proof of the strong near field interaction when f<0.07.

We implement the previously computed thermal conductance to determine the effective thermal conductivity of a nanoparticles distribution. We highlight the existence of an effective thermal

45 conductivity - i.e. linearity between flux and temperature difference –and the dependence to volume fraction. We will show that the Near-Field heat transfer channel is indeed significant and appear at volume fractions below the one corresponding to the mechanical percolation.

**2. The Physical Model**

50 We propose a statistical calculation of the thermal conductivity in nanocomposites due to near-field radiation based on the following assumptions:

(i) The contribution of the near-field heat transfer in nanocomposites can be derived from thermal interactions between pairs of NPs. The absence of 'collective' effect is due to the rapid decay of the thermal conductance between two NPs ($d^{-6}$, $d^{-20}$) and the reasonably low volume fractions (<10%);

55 (ii) The thermal interactions between NPs in composite media are the same as in vacuum. We argue that near-field transfer occurs at a well defined Terra Hertz frequencies in optically or electronically active materials. At those frequencies, the index of conventional liquids is near to one. We assume that the involved matrixes will neither absorb at those frequencies. Also, the separation distance d is much smaller than the extinction length at any frequency.

60 The system consists of an ensemble of NPs with a uniform spatial distribution. The number of NPs in a given volume – a parallelepiped - prescribes the volume fraction. As illustrated in Figure 1, the opposite walls of the box have different temperatures $T_1$ and $T_2$ along the direction z. In the transverse directions, adiabatic boundary conditions are applied. The heat flux crossing a given section referenced by a z coordinate can be expressed as:

65 $$\varphi(z) = \frac{1}{2} \sum_{i,j=1}^{N} \frac{\Gamma}{r_{ij}^n} (T_i - T_j)\left[H(z-z_i) - H(z-z_j)\right], \qquad (1)$$

where the heat flux due to a pair of NPs is expressed as a function of the thermal conductance $G^{NF} = (\Gamma/r_{ij}^n)$.[4] The quantity $r_{ij}$ is the separation distance between NPs j and i and n equals to -6 in the range of the dipole-dipole approximation (d>4a) and n equals to -20 when d<4a. H refers to the Heaviside function. The term into brackets cancels the contributions of pairs that are both on the left

70 side or both on the right side of the cross section as explained in Figure 1. The one half factor takes into account the identical contributions of pairs ij and ji. $\Gamma$ is material and radius dependent and is defined as:

$$\Gamma = \frac{3}{4\pi^3} \int_0^\infty \alpha''^2(\omega) \frac{d\Theta(\omega)}{dT} d\omega, \qquad (2)$$

where $\alpha''$ is the imaginary part of the particle polarisability. This quantity is currently defined as

$\alpha'' = \frac{4\pi a^3}{3} \text{Im}\left(\frac{\varepsilon_{NP} - \varepsilon_{med}}{\varepsilon_{NP} + 2\varepsilon_{med}}\right)$. $\varepsilon$ is the dielectric constant of the NP (index NP) or of the medium

(index med). $\Theta(\omega) = \hbar\omega / \left(e^{-\frac{\hbar\omega}{k_B T}} - 1\right)$ is the energy in the frequency interval [$\omega$, $\omega+d\omega$] where $k_B$ represents the Boltzmann constant. a is the NP radius. In the following, we will however consider the value of $\Gamma$ fitted from the direct computation of $G^{NF}$.[4]

The total heat flux crossing the nanoparticle distribution, $\Phi$, is then defined as:

$$\Phi = \frac{1}{L_Z} \int_0^L \varphi(z) dz = \frac{1}{2L_Z} \sum_{i,j} \frac{\Gamma}{r_{ij}^n} (T_i - T_j)(z_j - z_i), \qquad (3)$$

where $L_Z$ is the box size in the z-direction.

### 3. The numerical calculation

We performed a numerical calculation of $\Phi$ based on Equation (3). Several ensembles of N NPs positions are randomly generated based on an uniform distribution probability. Each NP is assimilated to a thermal node with a given temperature $T_i$. Each node is connected to its neighbours according to the near-field thermal conductance $G^{NF}$. The temperatures of the particles situated in the system boundaries were set to the corresponding thermostat temperatures. The flux balance equation is defined for each node i as follows:

$$\sum_{\substack{j=1,N \\ j \neq i}} \varphi_{ij} = \sum_{\substack{j=1,N \\ j \neq i}} \frac{\Gamma}{r_{ij}^n} (T_i - T_j) = 0. \qquad (3)$$

This system of N coupled equations is solved by an iterative technique because the thermal interactions are strongly non-linear and because of the high values of N. A matrix based calculation would be too expensive and not accurate. The calculations include silica NPs of radius a=1nm. Φ is proportional to the radius to the power 6 because Γ is linearly dependent to the square of the NP polarizability. However, when keeping the same volume fraction, the larger the radius a, the larger the distance $r_{ij}$. We therefore do not expect a drastic increase of the flux with the NP size. Contrarily, the variations of Γ due to a change in the material are of several orders of magnitude. Polar materials and metals are good candidates for the thermal conductivity enhancement because they carry confined phonon-polaritons and plasmons.

The thermostatted regions have temperatures $T_1$=120°C and $T_2$=50°C. The results were not affected by a change in the thermostats properties such as temperature and number of particles in the thermostatted regions.

Figure 3 shows that the temperature profile can indeed be approximated by a linear law although temperature steps can be observed. We presume that this effect is due to a local NP rarefaction which should be removed by simulating a larger number of NPs. We have checked that this profile is not dependent on the initial conditions by starting the iterative scheme from various profiles (flat, steps, linear).

The effective thermal conductivity λ was derived as the ratio between the heat flux Φ and the temperature difference ($T_1$-$T_2$). Ensembles of N=1000 NPs were simulated. In those conditions, the fluctuation amplitude of λ is about 20%. We therefore perform twenty runs to obtain one ensemble average with less than 5% of inaccuracy. As reported in Figure 3 when f=0.01, the dependence of λ to the transverse and longitudinal sizes of the box are neglectible when $L_z$>200 nm and $L_x$, $L_y$>40 nm. This allows us to set the box dimensions for volume fractions f>0.01.

## 4. Results and Discussion

The effective thermal conductivity λ is reported as a function of the volume fraction in Figure 4. Those data were obtained from 200 calculations of temperature distributions. Those results are quite general from the qualitative point of view because changing the NP material or radius will only result in the change of the value of the parameter Γ. The thermal conductivity will then remain proportional to Γ.

Regarding to the quantitative aspects, the near-field contribution reaches the order of magnitude of the liquid conduction when the volume fraction approaches 3% and approaches the bulk $SiO_2$ contribution when f=0.08.

Let us estimate the percolation thresholds to approach the qualitative aspect. The literature indicates that a 2-dimensionnal disk distribution is characterized by a percolation threshold (apparition of the first infinite cluster) at the critical volume fraction f=0.5.[6] Considering that the particles are equally spaced and occupying each a square (in 2D) or a cube (in 3D) allows us to write $f_{2D} = \dfrac{\pi a^2}{d^2}$ and $f_{3D} = \dfrac{4\pi a^3}{3d^3}$. Deriving the ratio a/d from $f_{2D}$=0.5 leads to the critical volume fraction of $f_{3D} = \dfrac{4\pi}{3}\left(\dfrac{1}{2\pi}\right)^{3/2}$ =0.266. Now considering that the Near-Field interaction doubles the effective radius,[4] we define a new percolation value $f_{3D}^{NF} = \dfrac{4\pi}{3d^3}\left(\dfrac{a}{2}\right)^3$ =0.266/8=0.033. This calculation is an approximation because the transition to Near-Field contact is continuous while the percolation threshold is estimated based on contact/no-contact states. However, this estimation quite well predicts the change in behaviour observed for the thermal conductivity in Figure 4. This percolation appears much earlier than the expected one. $f_{3D}^{NF}$ is also smaller than the volume fraction defined when the separation distance d=4a, i.e. at the Near-Field 'contact'. This Near-Field threshold is entirely due to the change of behaviour when the dipole-dipole approximation breaks down. It is therefore decoupled from the NP radius or material. The radius and polarizability will only affect the level of the radiation. If the thermal interaction is as intense as predicted by the MD calculations, those remarks could lead to simple experimental proofs of the

Near-Field radiation in NPs based composites with adequate volume fractions. For instance, studies on nanofluids have revealed an unexpected thermal conductivity increase at low volume fraction.[1] The NPs materials were oxides which are efficient media regarding to Near-Field interaction.

Note that the Near-Field interaction does not depend on the NP-matrix thermal contact. This discards the issues encountered in Carbone Nanotubes solutions. Increasing the thermal conductivity could finally be performed by including nano-objects with a high aspect ratio and adequate polarizabilities. In the case of embedded metallic particles in a dielectric media, the electrical insulation can be preserved while thermal transport is enhanced. This point is crucial for heat sinks in microelectronics.

## 5. Conclusion

We have shown that the near-field radiation between NPs might generate an early percolation in nano-object distributions. The onset of Near-Field percolation was shown to happen for the volume fraction f=0.033 which is a much lower value than the one obtained for the mechanical percolation. A clear experimental proof of the radiation contribution could be performed by measuring the thermal conductivity of materials or fluids with various volume fractions. Following our previous calculations performed with no interaction between the particle and the matrix, the effective thermal conductivity due to radiation reaches the level of the bulk materials.

**REFERENCES**

[1] S. Lee, S.U.S. Choi, S. Li, J.A. Eastman. Measuring thermal conductivity of fluids containing oxide nanoparticles. ASME J. Heat Transfer, vol. 121, 280, (**1999**).

[2] S. H. Munson-McGee. Estimation of the critical concentration in an anisotropic percolation network. Phys. Rev. B, vol. 43, 3331, (**1991**).

[3] P. Keblinski, S.R. Phillpot, S.U.S. Choi, J.A. Eastman. Mechanisms of heat flow in suspensions

of nano-sized particles (nanofluids). International Journal of Heat and

Mass Transfer, vol. 45, 833, (**2002**).

[4] N. Shenogina, S. Shenogin, L. Xue, and P. Keblinski. On the lack of thermal percolation in carbon nanotube composites. App. Phys. Lett., vol. 87, 133106, (**2005**).

[5] G. Domingues, S. Volz, K. Joulain, J.J. Greffet. Heat Transfer between Two Nanoparticles Through Near Field Interaction. Phys. Rev. Lett., vol. 94, 85901, (**2005**).

[6]D.W. Hermann and D. Stauffer, Z. Phys. B, vol. 40, 133, (**1980**).

**CAPTIONS**

**Figure 1**: top: Schematic of the simulation box. A random distribution of nanoparticles is placed in a parallepiped with transverse adiabatic boundary conditions. Two thermostatted zones at temperature $T_1$=120°C and $T_2$=50°C are driving the heat flux across the medium. Bottom: Schematic of the local heat flux calculation of $\phi(z)$ (Equation (1)). The heat flux across a section (vertical dashed line) is computed as the sum of all the pair interactions. Only the links crossing the section are taken into account.

**Figure 2** : Effective thermal conductivity $\lambda$ of the nanocomposite versus the length of the simulation box Lz. When Lz>200nm, the thermal conductivity does not vary significantly. The insert shows the dependence of $\lambda$ versus the transverse length of the simulation box. When Lx>40nm, the value of l becomes reliable. The NP radius a=1nm and the volume fraction f=0.01.

**Figure 3** : Top: temperature profile in the nanocomposite obtained from the numerical calculation. Each data point represents a NP. The NP radius a is set to 1nm and the volume fraction f=0.01. A linear dependence is observed except in a few intervals were steps are observed. Bottom: 3-dimensionnal representation of the NP distribution. The colour indicates the temperature, spatial units are nanometers.

**Figure 4** : Effective thermal conductivity $\lambda$ of the nanocomposite against volume fraction. The numerical results (diamonds) are compared to the bulk $SiO_2$ thermal conductivity (black continuous

line) and to a hundredth of the water thermal conductivity (grey continuous line). The Near-Field and mechanical percolation thresholds are indicated as vertical dashed lines. The threshold between the $d^{-6}$ and $d^{-20}$ (n=-6 and n=-20) behaviors for $G^{NF}$ is also indicated. The thermal conductivity due to the mechanical percolation $\lambda=\lambda_0\,(f-f_c)^2$ is reported (thick line).

Figure 1 :

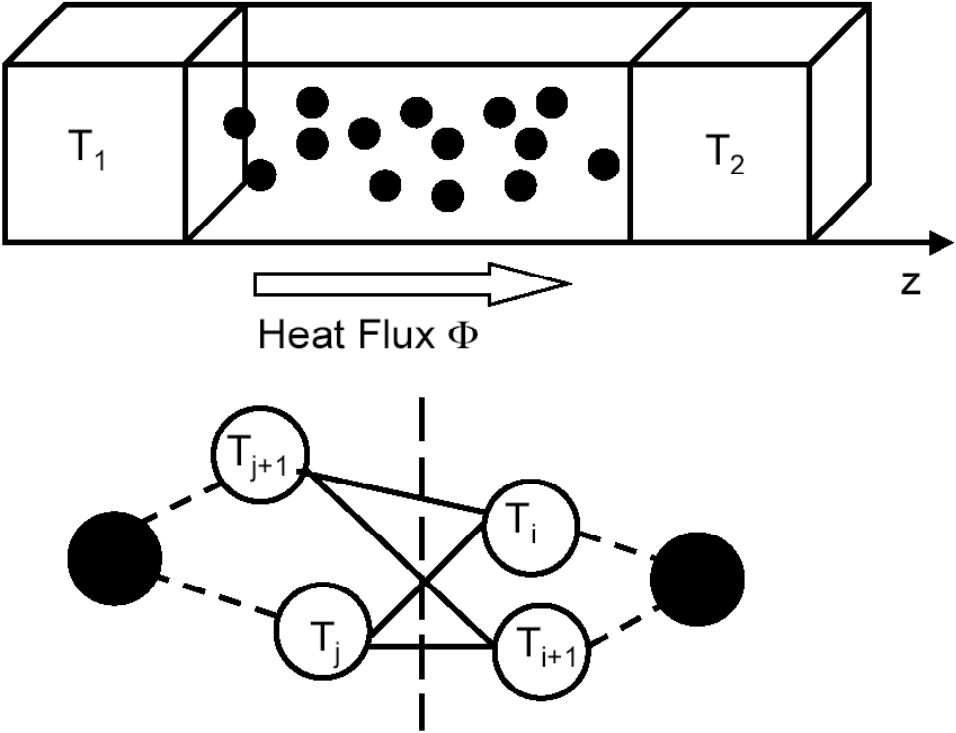

Figure 2 :

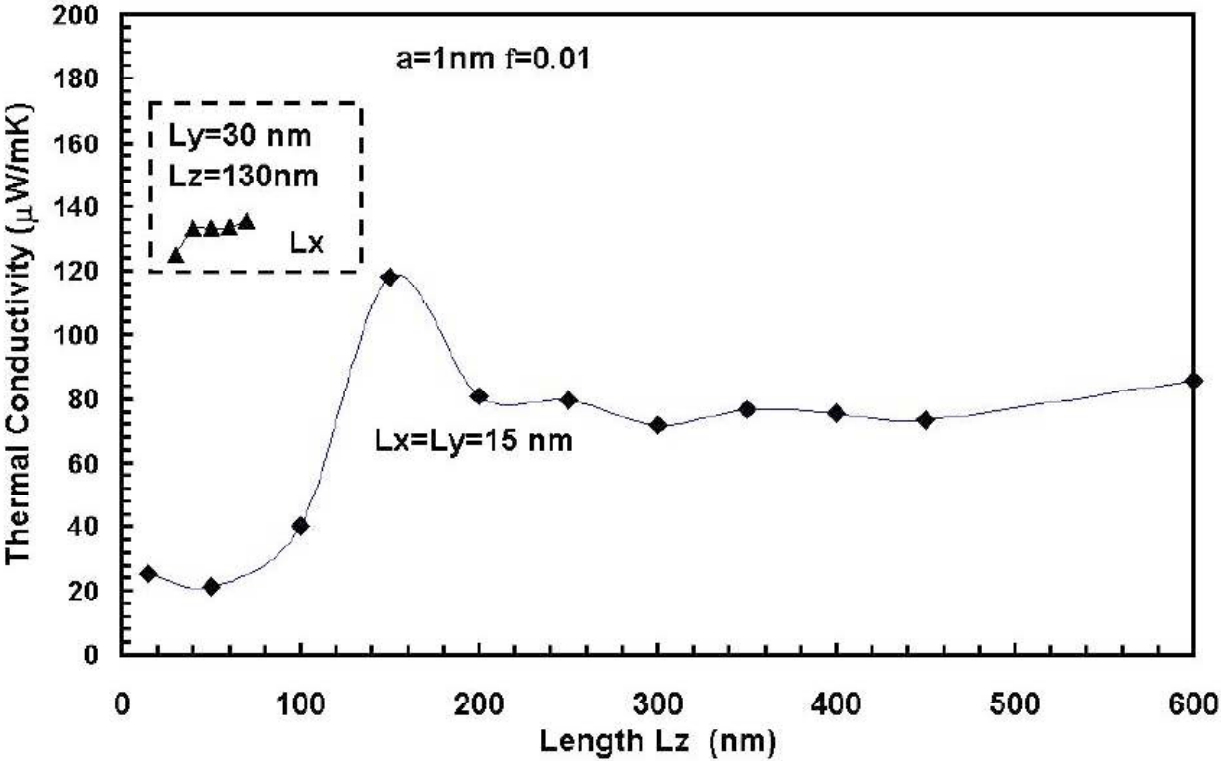

200

Figure 3 :

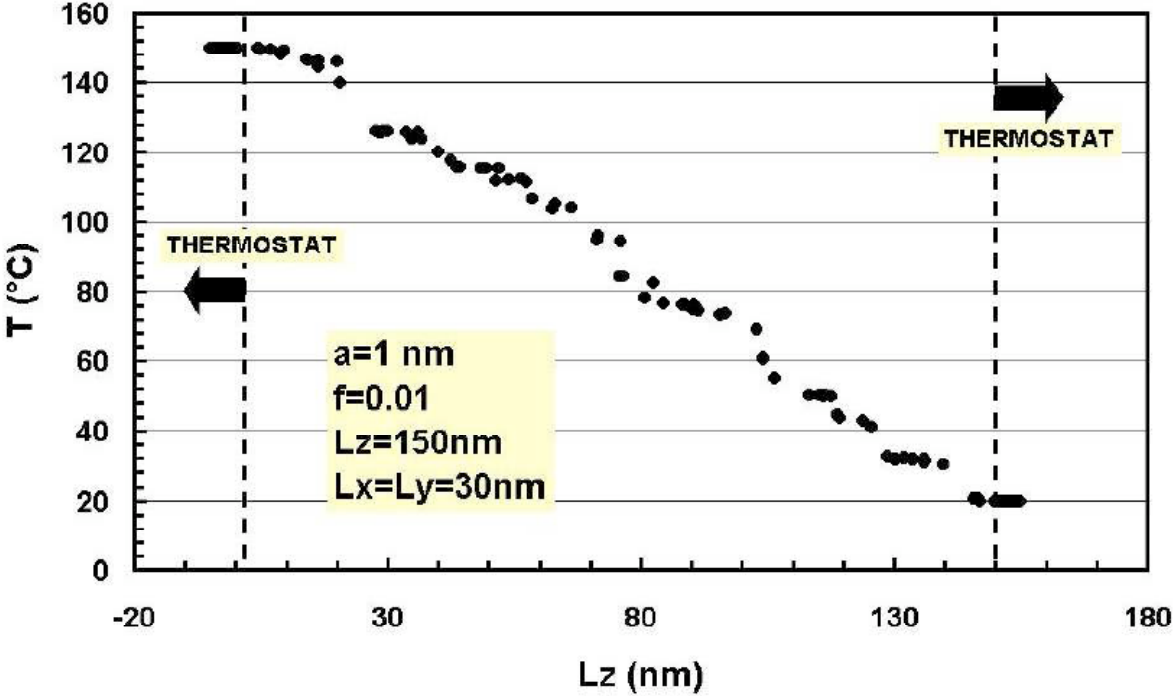

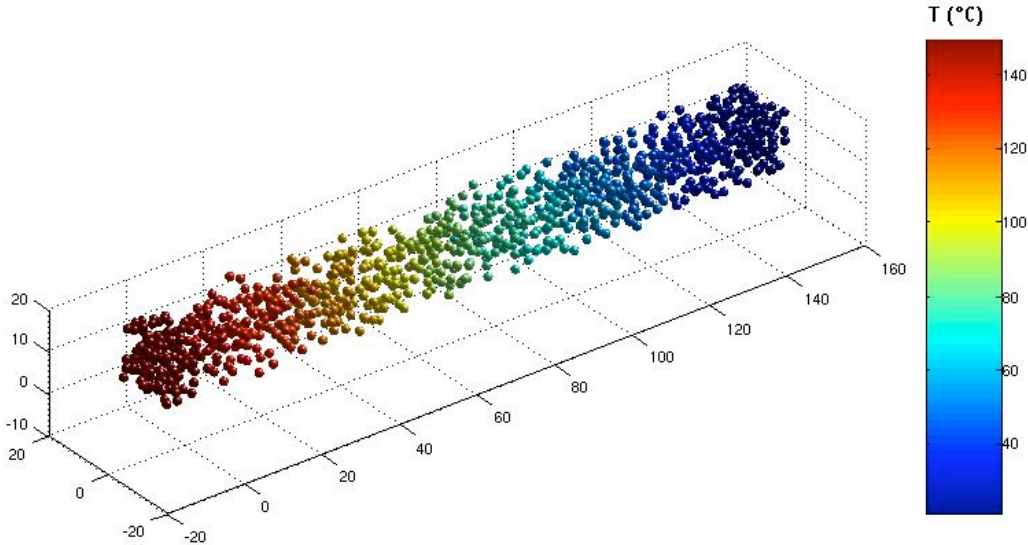

205

Figure 4 :

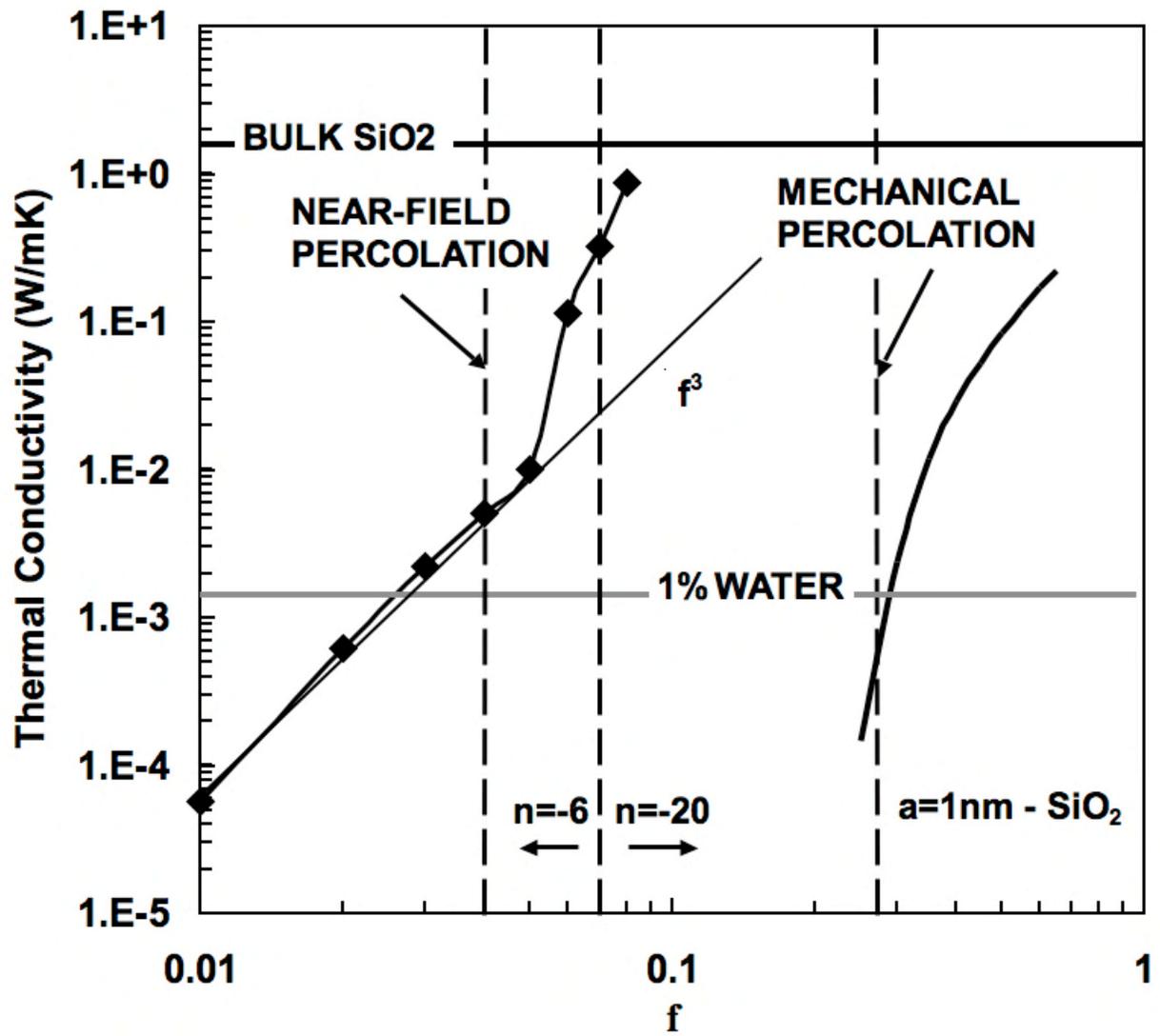